\documentclass[fleqn,usenatbib]{mnras}

\usepackage{newtxtext,newtxmath}

\usepackage[T1]{fontenc}

\DeclareRobustCommand{\VAN}[3]{#2}
\let\VANthebibliography\thebibliography
\def\thebibliography{\DeclareRobustCommand{\VAN}[3]{##3}\VANthebibliography}

\usepackage{graphicx}	
\usepackage{amsmath}	
\usepackage[flushleft]{threeparttable}
\usepackage{multirow}

\title[HD\,189733b NUV transit]{The near-UV transit of HD\,189733b with the XMM-Newton Optical Monitor}

\author[G. W. King et al.]{George W. King,$^{1,2,3}$\thanks{E-mail: kinggw@umich.edu}
L\'{i}a Corrales,$^1$
Peter J. Wheatley,$^{2,3}$
Panayotis Lavvas,$^4$
Maria E. Steinrueck,$^5$
\newauthor
Vincent Bourrier,$^6$
David Ehrenreich,$^6$
Alain Lecavelier des Etangs,$^7$
and Tom Louden$^{2,3}$
\\
$^{1}$Department of Astronomy, University of Michigan, Ann Arbor, MI 48109, USA\\
$^{2}$Department of Physics, University of Warwick, Gibbet Hill Road, Coventry, CV4 7AL, UK\\
$^{3}$Centre for Exoplanets and Habitability, University of Warwick, Gibbet Hill Road, Coventry, CV4 7AL, UK\\
$^{4}$Groupe de Spectrométrie Moléculaire et Atmosphérique, UMR CNRS 7331, Université Reims Champagne Ardenne, France\\
$^{5}$Lunar and Planetary Laboratory, University of Arizona, Tucson, AZ, 85721, USA\\
$^{6}$Observatoire de l'Universit\'{e} de Gen\`{e}ve, 51 chemin des Maillettes, 1290 Sauverny, Switzerland\\
$^{7}$Institut d'Astrophysique de Paris, UMR7095 CNRS, Universit\'{e} Pierre \& Marie Curie, 98bis boulevard Arago, 75014 Paris, France
}

\date{Accepted XXX. Received YYY; in original form ZZZ}

\pubyear{2021}

\begin{document}
\label{firstpage}
\pagerange{\pageref{firstpage}--\pageref{lastpage}}
\maketitle

\begin{abstract}
We present analysis of \textit{XMM-Newton} Optical Monitor observations in the near-ultraviolet of HD\,189733, covering twenty primary transits of its hot Jupiter planet. The transit is clearly detected with both the UVW2 and UVM2 filters, and our fits to the data reveal transit depths in agreement with that observed optically. The measured depths correspond to radii of $1.059^{+0.046}_{-0.050}$ and $0.94^{+0.15}_{-0.17}$ times the optically-measured radius (1.187\,R$_{\rm J}$ at 4950\,\AA) in the UVW2 and UVM2 bandpasses, respectively. We also find no statistically significant variation in the transit depth across the 8 year baseline of the observations. We rule out extended broadband absorption towards or beyond the Roche lobe at the wavelengths investigated, although observations with higher spectral resolution are required to determine if absorption out to those distances from the planet is present in individual near-UV lines.
\end{abstract}

\begin{keywords}
planets and satellites: individual: HD\,189733b -- ultraviolet: planetary systems -- planets and satellites: atmospheres
\end{keywords}



\section{Introduction}
One of the key aspects in characterising discovered exoplanets is investigating their atmospheric composition. Transiting planets are particularly good targets for this as the apparent transit depth varies as a function of wavelength. This variation is driven by the bulk elemental composition, molecular, and particulate species present in any atmosphere maintained by the planet. This method of ``transmission spectroscopy" has been widely applied to discover numerous species in the atmospheres of hot Jupiters \citep[e.g.][]{Sing2016}. Attempts to explore much smaller super-Earths \citep[e.g.][]{Kreidberg2014,Edwards2020} and even Earth-sized planets \citep[e.g.][]{deWit2018} are now being made.

The near-ultraviolet (NUV) is an intriguing wavelength range to target for transmission spectroscopy. There are numerous lines of metallic species that may be observable in absorption \citep{Lothringer2020}, including neutral and singly ionised Fe and Mg. Such absorption can arise from materials in an extended or escaping atmosphere \citep{Fossati2010,Haswell2012,Sing2019,Cubillos2020}. As such, NUV transits could provide insight into mass loss from exoplanets, complementing observations at other wavelengths such as Ly\,$\alpha$ \citep[e.g.][]{VM2003,LDE2012,Ehrenreich2015}. Additionally, optically-measured Rayleigh scattering slopes may extend into the NUV.

The NUV has been largely underexplored. Ground-based measurements can only be made down to about 3000\,\AA. Furthermore, where observations have been taken, it can be challenging to interpret the results due to the relatively small number of strong, unblended lines, as compared to other wavelengths more commonly used for exoplanet investigations. From space, the \textit{Hubble Space Telescope} (\textit{HST}) is capable of observing in the NUV with its COS, STIS, and WFC3/UVIS instruments \citep{Fossati2010,Haswell2012,VM2013,Sing2019,Wakeford2020,Cubillos2020}, though the majority of transmission spectroscopy experiments performed with HST have focused on the optical, near-infrared, and some FUV \citep[e.g.][]{Kreidberg2014,Ehrenreich2015,Sing2016}.

In the last few years, both the \textit{XMM-Newton} Optical Monitor (OM) and \textit{Swift} Ultraviolet/Optical Telescope (UVOT) have been used to detect planetary transits in the NUV. While both facilities are primarily used to make high-energy observations, the 30\,cm diameter OM and UVOT observe simultaneously with the various X-ray and gamma-rays telescopes, and both have a range of broad-band filters in the optical and NUV. \citet{W80} made the first detection of a NUV transit with the OM, using the UVW1 filter to observe a transit of WASP-80b. While consistent with the optically measured transit depth, there was a hint of it being shallower, a result also suggested by an earlier ground-based U-band observation \citep{Turner2017}. The first detection using UVOT was reported for WASP-121b by \citet{Salz2019}, with the measured transit depth deeper at the 2-$\sigma$ level than in the optical. 

Here, we analyse data taken with the \textit{XMM-Newton} OM across twenty primary transits of the prototypical transiting hot Jupiter HD\,189733b, wherein we detect the transit in two different broadband NUV filters.

\subsection{The HD 189733 system}
The discovery of HD\,189733b was reported by \citet{Bouchy2005}, and it remains the closest transiting hot Jupiter to Earth, orbiting a relatively active K1 dwarf at a distance of just $19.775\pm0.013$\,pc~\citep{GaiaDR2}. This fact has led to HD\,189733b being one of the most popular targets for both theoretical studies, and follow-up observations to characterise its atmosphere. We give the parameters of the system adopted in this study in Table~\ref{tab:189param}.

At optical wavelengths, the transmission spectrum of HD\,189733b shows a steep-gradient slope \citep[e.g.][]{Pont2008,Sing2011,Gibson2012}. Excess absorption has been observed at the \ion{Na}{I} doublet \citep[e.g.][]{Redfield2008,Huitson2012,Wyttenbach2015,Louden2015,Khalafinejad2017}, and there is evidence for a \ion{K}{I} absorption feature \citep{Pont2013,Keles2019,Keles2020}. In the near-infrared, water \citep{Birkby2013,McCullough2014,Brogi2016,Brogi2018} and CO \citep{deKok2013,Rodler2013,Brogi2016} have both been detected. The subdued amplitude of both the water and wings of the sodium line features, together with the steep optical slope, point towards the presence of high-altitude aerosols in the atmosphere \citep{Pont2013,Sing2016}. The planet is also one of an increasing number to have a helium excess measured in the metastable 10830\,\AA\ triplet \citep{Salz2018,Guilluy2020}.

HD\,189733b's atmosphere has also been studied at shorter wavelengths, with Ly\,$\alpha$ transit observations showing that \ion{H}{i} is moving beyond the Roche lobe and escaping the atmosphere \citep{LDE2010,LDE2012,Bourrier2013,Bourrier2020}. The transit depths at Ly\,$\alpha$ wavelengths of up to 15 per cent have been observed to be variable \citep{LDE2012}. Additionally, \citet{BenJaffel2013} measured a 6.4 per cent transit in \ion{O}{I}, and there is a time-variable absorption signature in the \ion{Si}{iii} line that could arise from a bow-shock formed ahead of the planet in its orbit \citep{Bourrier2013,Bourrier2020}. In X-rays \citet{Poppenhaeger2013} presented evidence of the X-ray transit possibly being as deep as 8 per cent. Our analysis of the simultaneously taken X-ray data will be published separately (Wheatley et al., in prep; King et al., in prep). Taken together, these observations show that the XUV heating of HD\,189733b has led to its atmosphere being extended, and in at least the case of \ion{H}{i}, escaping. As is thought to be the case for almost all hot Jupiters \citep[e.g.][]{Owen2012,B+LDE2013}, the rate of escape of material becoming unbound to the planet is not high enough to significantly change the planet's structure.

\begin{table}
  \caption{Adopted stellar and planetary parameters for HD\,189733(b).}
  \label{tab:189param}
  \centering
  \begin{threeparttable}
  \begin{tabular}{l l c l c}
    \hline
    Parameter & Symbol & Value & Unit & Ref. \\
    \hline
    Stellar mass & $M_*$            & $0.823\pm0.029$  & M$_\odot$ & 1\\
    Stellar radius & $R_*$ & $0.780^{+0.017}_{-0.024}$ & R$_\odot$ & 2\\
    Planet to star rad. & $R_{\rm p}/R_*$ & $0.15641\pm0.00010$$^\dagger$ & & 3\\
    Orbital period & $P_{\rm orb}$ & 2.218575200(77) & d & 4\\
    Transit centre & $T_0$ & 2453955.5255511(88) & BJD$_{\rm TDB}$ & 4\\
    Semi-maj. axis & \multirow{2}{*}{$a/R_*$} & \multirow{2}{*}{$8.863\pm0.020$} &  & \multirow{2}{*}{5}\\
    to star radius & & & & \\
    Eccentricity & $e$                         & 0  &               & 6\\
    Orbital inclin. & $i$              & $85.710\pm0.024$  & $^\circ$ & 5\\
    \hline
\end{tabular}
\begin{tablenotes}
\item References: (1) \citet{Triaud2009}; (2) \citet{GaiaDR2}; (3) \citet{Sing2011}; (4) \citet{Baluev2015}; (5) \citet{Agol2010}; (6) \citet{Bouchy2005}.
\item $^\dagger$ As measured at 4950\,\AA.
\end{tablenotes}
\end{threeparttable}
\end{table}

\section{Observations \& data reduction}
\textit{XMM-Newton} has observed HD\,189733b on 25 separate occasions from 2007 through 2015, with twenty of these covering a primary transit. In Table~\ref{tab:Obs}, we present details of these twenty transit observations (the PI in each case was P. Wheatley).

Our analysis here focuses on the data taken with the Optical Monitor, a 30\,cm aperture telescope with a photon-counting instrument at Cassegrain focus, operating in the visual and NUV \citep{Mason2001}. The observations presented in the work exploited the rare ultraviolet capabilities of the OM, using the UVW1, UVW2, and UVM2 filters\footnote{For more information about the filters of the OM, see the \textit{XMM-Newton} Users Handbook: \url{http://xmm-tools.cosmos.esa.int/external/xmm_user_support/documentation/uhb/omfilters.html}}. All twenty of these observations were taken in imaging mode, along with a single, small fast mode window to capture the light from HD\,189733 at 11\,ms resolution. 

As per their definition in Table~\ref{tab:Obs}, observations 1 and 2 employed the UVM2 (effective wavelength = 2910\,\AA; width\footnote{The width of a filter with a constant transmission equal to that at the effective wavelength, and which has the same effective area as the real OM filter in question.} = 830\,\AA) and UVW1 (2310\,\AA; 480\,\AA) filters, respectively, while the other 18 all used the UVW2 (2120\,\AA; 500\,\AA) filter. The observation 2 data were rendered unusable by the brightness of the source resulting in excessive coincidence losses that could not be corrected. We do not analyse or discuss this observation any further.
%
The UVM2 filter choice for observation 1 lead to a count rate of 9.4\,s$^{-1}$, and the 18 observations that employed the UVW2 filter had an average count rate of 4.9\,s$^{-1}$.

We reduced the data using the standard software \textsc{omichain} and \textsc{omfchain} within the Scientific Analysis System, for the image and fast mode data, respectively. Although the fast mode data is captured at 11\,ms resolution, in practice, the standard reduction pipeline produces a time-series file at 10\,s resolution by default, a setting which we do not alter. Following the running of the chains, we corrected each observation's photometric data using the procedure as described in \citet{W80}, wherein we correct the fast mode time series using the image mode data.

\begin{table*}
\centering
\caption{Table detailing the twenty \textit{XMM-Newton} observations of HD\,189733 analysed in this paper.}
\label{tab:Obs}
\begin{tabular}{rcccccc}
\hline
No. & ObsID & Start Time & End Time & Start -- Stop & Exp. Time & Filter\\
    &       & (BJD$_{\rm TDB}$) & (BJD$_{\rm TDB}$) & Phase &  (ks) & \\
\hline
1  & 0506070201 & 2007-04-17T14:16 & 2007-04-18T04:36 & 0.8431 -- 1.1121 & 44.0 & UVM2 \\
2  & 0692290201 & 2013-05-09T20:26 & 2013-05-10T07:02 & 0.8965 -- 1.0958 & 33.5 & UVW1 \\
3  & 0692290301 & 2013-11-03T08:21 & 2013-11-03T17:42 & 0.9015 -- 1.0768 & 30.7 & UVW2 \\
4  & 0692290401 & 2013-11-21T01:26 & 2013-11-21T12:38 & 0.8847 -- 1.0950 & 37.4 & UVW2 \\
5  & 0744980201 & 2014-04-05T05:33 & 2014-04-05T17:24 & 0.8118 -- 1.0346 & 39.8 & UVW2 \\
6  & 0744980301 & 2014-05-02T01:50 & 2014-05-02T07:40 & 0.9120 -- 1.0217 & 18.8 & UVW2 \\
7  & 0744980401 & 2014-05-13T02:23 & 2014-05-13T12:31 & 0.8804 -- 1.0709 & 31.8 & UVW2 \\
8  & 0744980501 & 2014-05-15T10:24 & 2014-05-15T18:49 & 0.9327 -- 1.0908 & 25.6 & UVW2 \\
9  & 0744980601 & 2014-05-17T14:48 & 2014-05-17T23:13 & 0.9168 -- 1.0749 & 27.4 & UVW2 \\
10 & 0744980801 & 2014-10-17T16:36 & 2014-10-18T02:24 & 0.9136 -- 1.0978 & 28.8 & UVW2 \\
11 & 0744980901 & 2014-10-19T21:06 & 2014-10-20T06:11 & 0.8996 -- 1.0702 & 28.0 & UVW2 \\
12 & 0744981001 & 2014-10-22T02:06 & 2014-10-22T12:43 & 0.8952 -- 1.0945 & 33.5 & UVW2 \\
13 & 0744981101 & 2014-10-24T06:43 & 2014-10-24T17:05 & 0.8833 -- 1.0779 & 32.6 & UVW2 \\
14 & 0744981301 & 2014-11-08T20:44 & 2014-11-09T05:52 & 0.9075 -- 1.0792 & 28.2 & UVW2 \\
15 & 0744981201 & 2014-11-11T01:05 & 2014-11-11T12:50 & 0.8907 -- 1.1114 & 39.4 & UVW2 \\
16 & 0744981401 & 2014-11-13T07:13 & 2014-11-13T15:08 & 0.9075 -- 1.0562 & 25.6 & UVW2 \\
17 & 0744980701 & 2014-11-15T10:15 & 2014-11-15T20:07 & 0.8660 -- 1.0512 & 30.8 & UVW2 \\
18 & 0744981501 & 2015-04-13T03:05 & 2015-04-13T14:13 & 0.8914 -- 1.1006 & 35.4 & UVW2 \\
19 & 0744981601 & 2015-04-17T13:02 & 2015-04-17T22:57 & 0.8814 -- 1.0677 & 31.0 & UVW2 \\
20 & 0744981701 & 2015-04-19T19:34 & 2015-04-20T05:39 & 0.9054 -- 1.0948 & 31.6 & UVW2 \\
\hline
\end{tabular}
\end{table*}

\section{Data analysis \& results}

\subsection{UVW2 observations}
\label{ssec:UVW2}

\begin{figure}
\centering
 \includegraphics[width=\columnwidth]{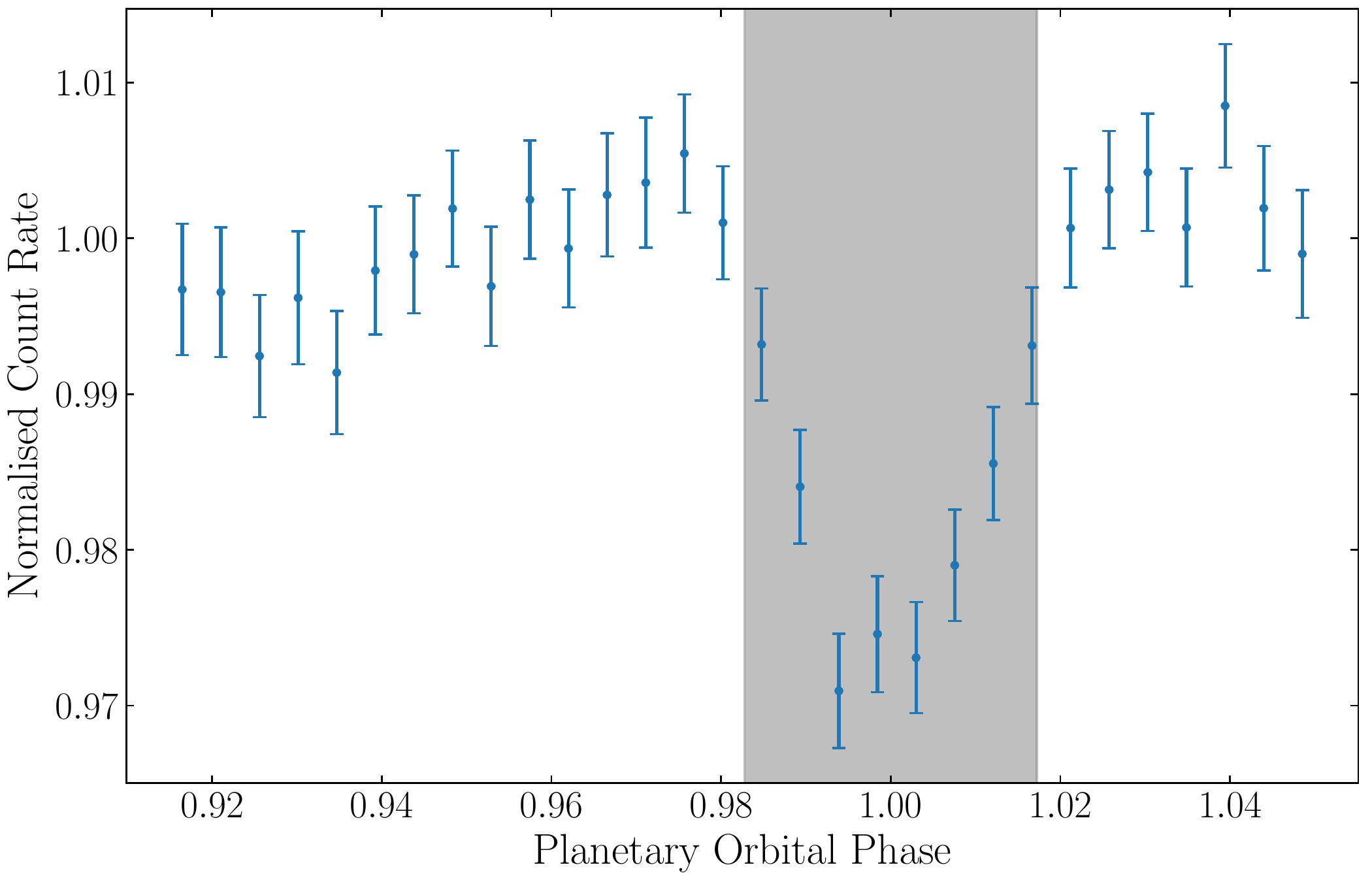}
 \caption{Binned, phase folded OM light curve for the 18 observations with UVW2. No correction has been made in this figure for out of transit trends. The shaded region is the full duration of the optical transit, with the phases plotted according to the ephemeris in Table~\ref{tab:189param}.}
 \label{fig:OMfoldRaw}
\end{figure}

The bulk of our analysis efforts focused on the 18 observations taken with the UVW2 filter. We restrict our analysis to the phases where there are data for at least 16 of the 18 observations: 0.9142--1.0508. Visual inspection of the binned, phase-folded light curve, shown in Fig.~\ref{fig:OMfoldRaw} revealed a clear transit detection at the expected time, according to the optical ephemeris. The light curve however shows structure in the out of transit data. By examining the individual observation light curves, we saw that this out of transit trend varied in shape from observation to observation.

We performed several MCMC fits to the time series data, binned to the reduction pipeline standard cadence of 10\,s, and with each observation normalised to its out of transit count rate. We used the \textsc{batman} \citep{Kreidberg2015} transit model with a quadratic limb darkening law, with Gaussian priors applied to the coefficients u$_1$ = $0.0594\pm0.0200$ and u$_2$ = $0.0160\pm0.0200$. These coefficients were determined by passing values of the effective area of the UVW2 filter as a function of wavelength through the \textsc{TabulatedFilter} function of Limb Darkening Toolkit \citep{Parviainen2015}. Gaussian priors were also placed on the system's inclination, $i$, and ratio of the semi-major axis and stellar radius, $a/R_*$, according to their values and uncertainties listed in Table~\ref{tab:189param}.

We accounted for the out of transit trends by multiplying the transit model by a quadratic, $a_jt^2 + b_jt + c_j$, allowing the coefficients to vary across the different observations, $j$. However, in order to help constrain the coefficients and avoid them going off to erroneous values, we initially performed a fit using a single quadratic that was the same for each observation, yielding $a=-0.997^{+0.697}_{-0.676}$, $b=2.02^{+1.33}_{-1.37}$, and $c=-0.016^{+0.676}_{-0.654}$. In our final fits, we placed uniform priors on the quadratic coefficients for each observation, forcing them to be within 2-$\sigma$ of this initial single quadratic fit. Additionally, at this same step, we ran a second, similar fit in which the only difference was that we allowed the mid-transit time $t_0$ to vary, in order to verify there was no offset in this value from the ephemeris present in the data. The best-fit $t_0$ was consistent to within 1-$\sigma$ of the optical ephemeris (in phase, $t_0 = 1.00057^{+0.00081}_{-0.00082}$). In all of our following analyses, we accordingly fixed $t_0$ according to the optically-measured ephemeris.

\begin{figure}
\centering
 \includegraphics[width=\columnwidth]{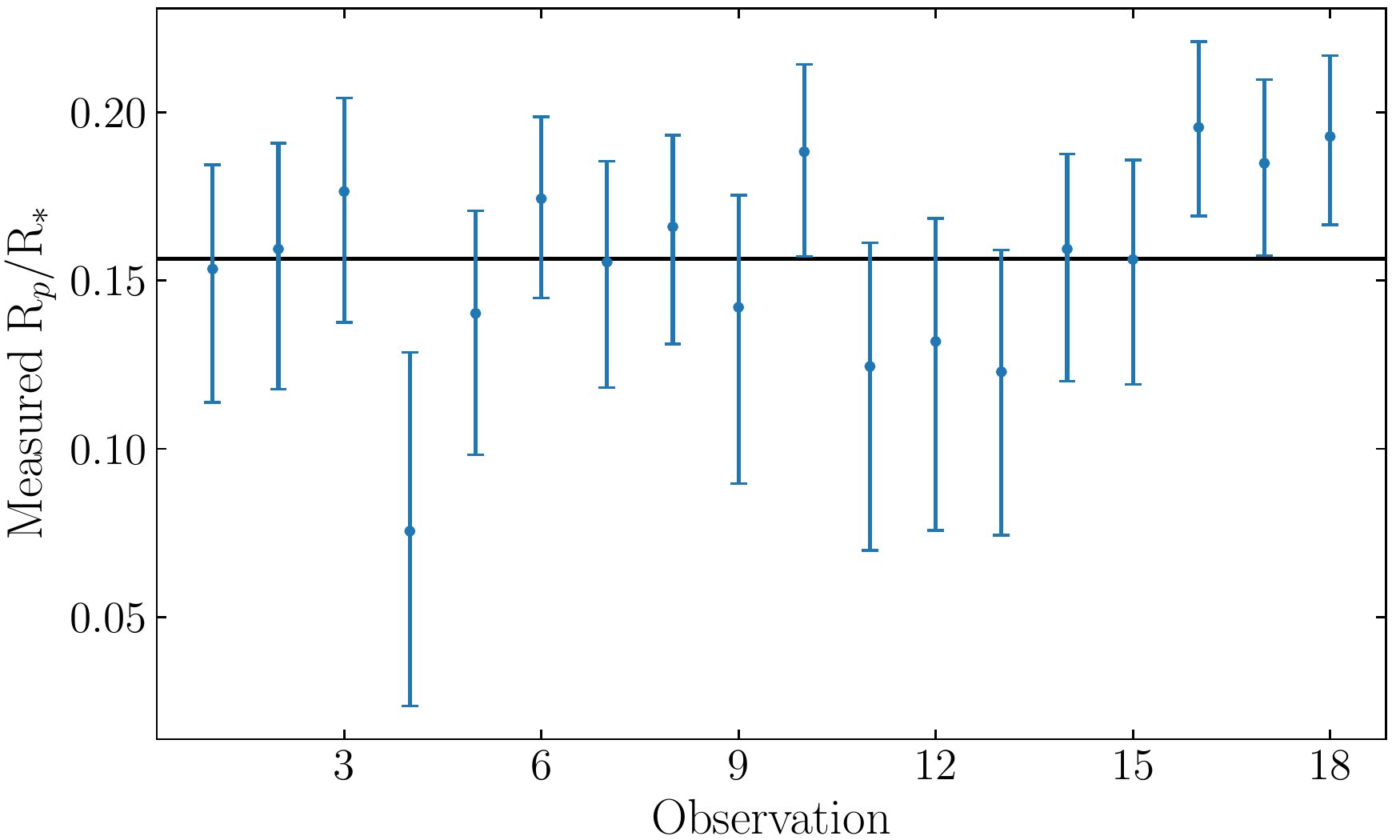}
 \caption{Measured value of the ratio of the planet and stellar radius, $R_{\rm p, UVW2}/R_*$, in each transit observation with the UVW2 filter on the Optical Monitor. The solid horizontal line depicts the measured value of $R_{\rm p}/R_*$ at optical wavelengths.}
 \label{fig:RpVary}
\end{figure}

Our three final fits investigated the measured planet radius in the UVW2 band ($R_{\rm p, UVW2}$). In the first of these, we allowed $R_{\rm p, UVW2}/R_*$ to change between all of the observations, applying a wide uniform prior, $0 < \dfrac{R_{\rm p, UVW2}}{R_*} < 1$, to prevent unphysical values. As we were interested in the relative changes of $R_{\rm p, UVW2}$ across each observation, in this fit we fixed the values of $i$, $a/R_*$, u$_1$, and u$_2$, as opposed to utilising the Gaussian priors. In Fig.~\ref{fig:RpVary}, we plot the measured $R_{\rm p, UVW2}/R_*$ as a function of observation number. This plot shows that the individual UVW2 transit depths are consistent with each other, and with the transit depths observed in the optical (see Table~\ref{tab:189param}). The mean and median of these 18 measurements are $R_{\rm p, UVW2}/R_* = 0.1555\pm0.0071$ and 0.159, respectively. In terms of $R_{\rm p, UVW2}/R_{\rm p, opt}$, where $R_{\rm p, opt}$ is the optically measured radius (at 4950\,\AA), these values are $0.994\pm0.045$ and 1.00. We statistically tested for variation of $R_{\rm p, UVW2}/R_*$ by comparing against a constant model equal to the mean of the 18 values, 0.15722. This gave $\chi^2_{\rm red} = 0.71$ and a p-value of 0.79, indicating that the values are consistent with being constant within the uncertainties.

\begin{figure}
\centering
 \includegraphics[width=\columnwidth]{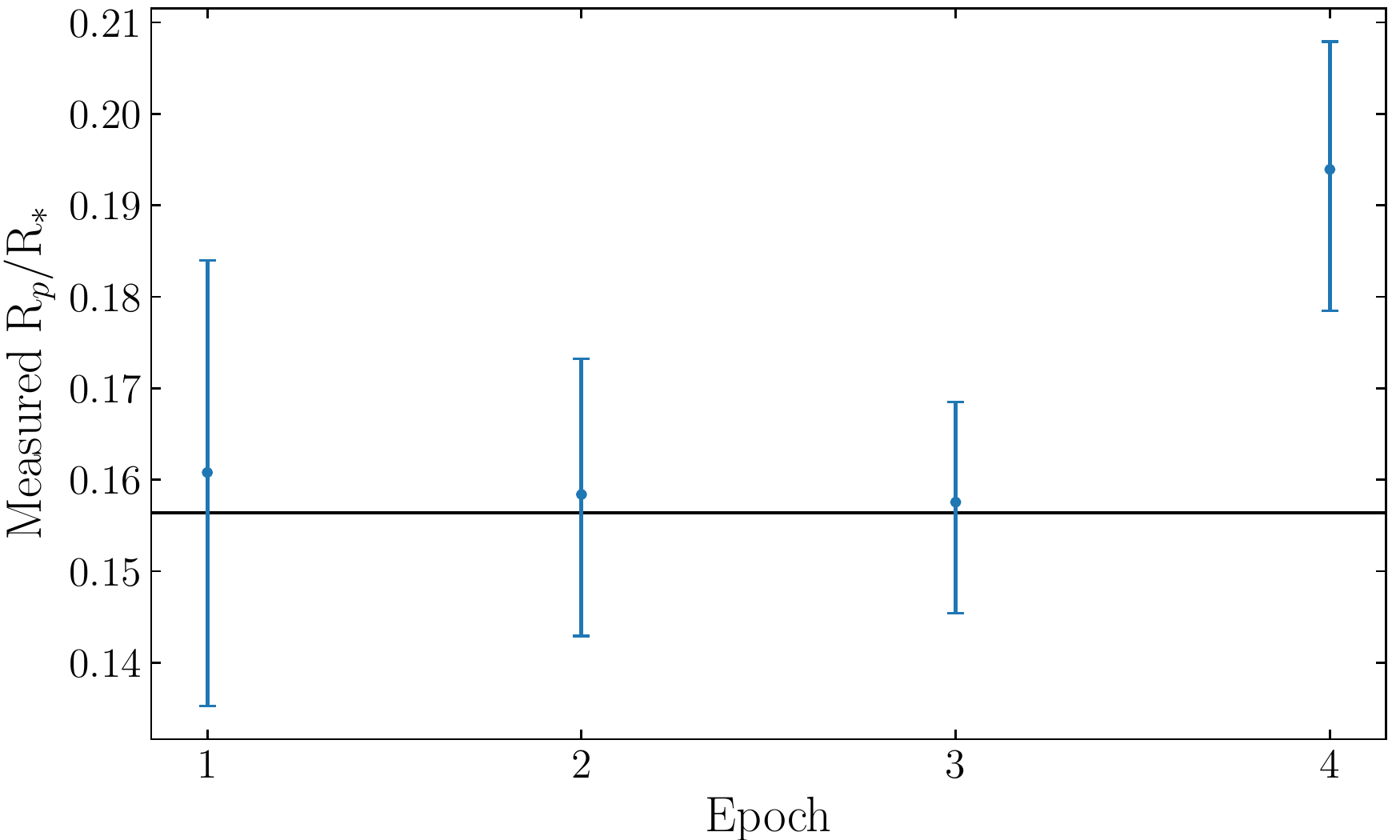}
 \caption{Measured value of the ratio of the planet and stellar radius, $R_{\rm p, UVW2}/R_*$, at each of the four defined epochs (see main text). The solid horizontal line depicts the measured value of $R_{\rm p}/R_*$ at optical wavelengths.}
 \label{fig:RpEpochs}
\end{figure}

In the second fit, we further investigated variation in the transit on longer timescales, by forcing $R_{\rm p, UVW2}/R_*$ to be the same for each observation within (but not between) the following defined epochs: "Autumn 2013" (Observations 3 and 4 in Table~\ref{tab:Obs}), "Spring 2014" (Observations 5 to 9), "Autumn 2014" (Observations 10 to 17), and "Spring 2015" (Observations 18 to 20). The same wide uniform prior as before was used for each of the four $R_{\rm p, UVW2}/R_*$ values, while $i$, $a/R_*$, u$_1$, and u$_2$ were again fixed for this fit. In Fig.~\ref{fig:RpEpochs}, we plot the measured $R_{\rm p, UVW2}/R_*$ for each of these four epochs. Although the final epoch for the Spring 2015 data shows a hint of a larger $R_{\rm p, UVW2}/R_*$ compared to the other three epochs, the four points are consistent with a constant model equal to the mean, with $\chi^2_{\rm red} = 1.47$ and a p-value of 0.22.

\begin{figure}
\centering
 \includegraphics[width=\columnwidth]{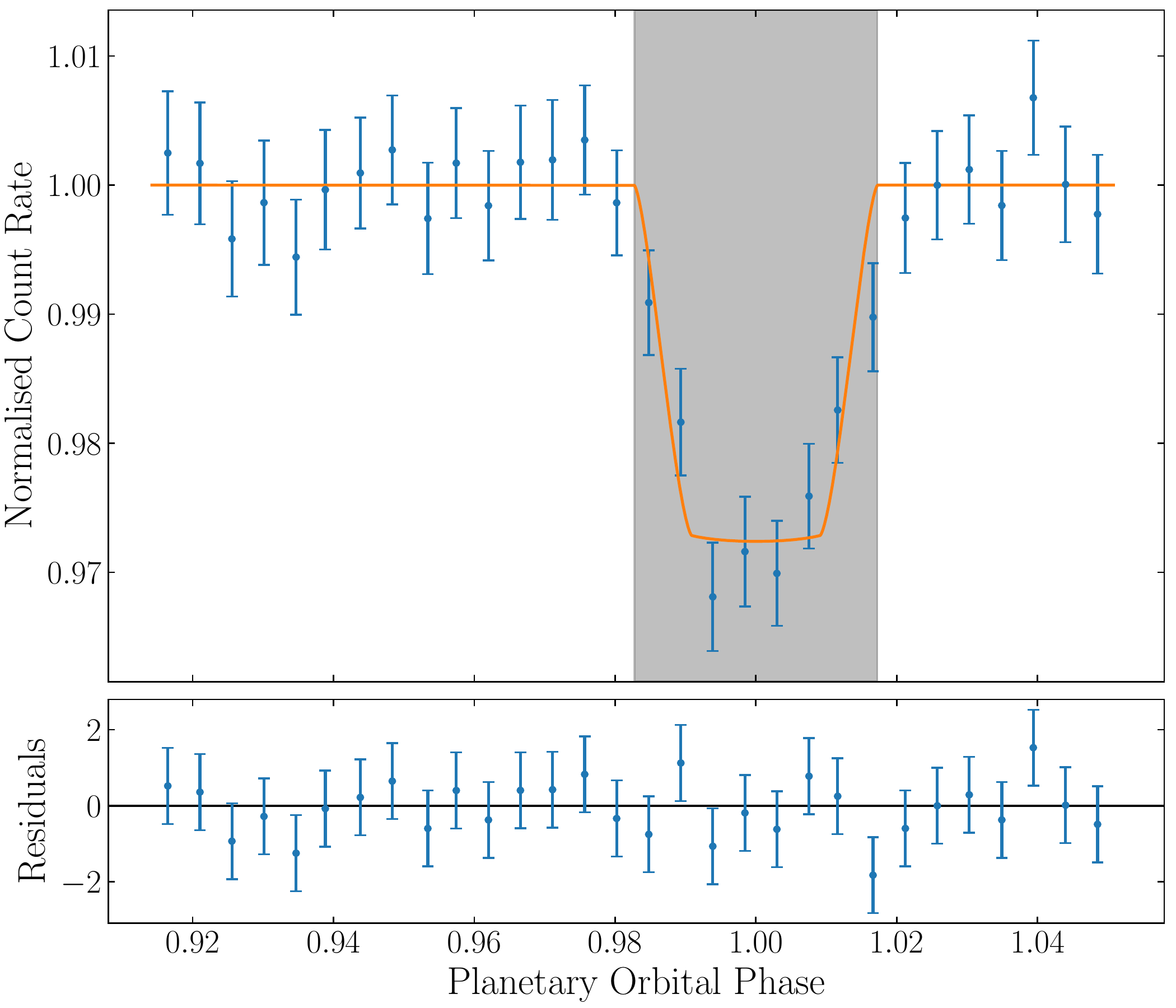}
 \caption{Top panel: binned, phase folded OM light curve for the 18 UVW2 observations, with the best fitting out of transit quadratic trends removed. Plotted in orange is the best fit model, where $R_{\rm p, UVW2}$ was forced to be constant across all 18 light curves. Bottom panel: residuals of the model.}
 \label{fig:om+model}
\end{figure}

Following these findings, we ran an MCMC wherein we forced $R_{\rm p, UVW2}/R_*$ to be the same across all 18 transits. The Gaussian priors on $i$, $a/R_*$, u$_1$, and u$_2$ were restored. In Fig.~\ref{fig:om+model}, we plot the phase folded light curve with the best fit model from this MCMC plotted over the top. In this plot, the best-fitting out-of-transit trends have been removed from each observation's light curve before phase folding and binning. Parameter details for this MCMC run are given in Table~\ref{tab:MCMC}. Most notably, this run gives a best-fitting value of $R_{\rm p, UVW2} = 1.059^{+0.046}_{-0.050}$\,R$_{\rm opt}$, a value which is consistent with the optical radius of the planet to just outside 1-$\sigma$. 

\begin{table}
\centering
\caption{Parameters used for and obtained from the final, best MCMC run for each OM filter's observations. In the case of UWV2, this is the fit where $R_{\rm p, UVW2}/R_*$ was forced to be the same across all observations.}
\label{tab:MCMC}
\begin{tabular}{lccc}
\hline
Parameter & \multicolumn{2}{c}{Value} & Unit \\
 & UVW2 & UVM2 \\
\hline
\multicolumn{4}{c}{\textit{Gaussian priors}} \\[0.1cm]
u$_1$           & $0.0594\pm0.0200$             & $0.0618\pm0.0250$\\
u$_2$           & $0.0160\pm0.0200$             & $0.0204\pm0.0250$\\
$a/R_*$          & \multicolumn{2}{c}{$8.863\pm0.020$}\\
$i$             & \multicolumn{2}{c}{$85.710\pm0.024$}                      & $^\circ$\\
\hline
\multicolumn{4}{c}{\textit{Fixed Value}} \\[0.1cm]
$t_0$           & \multicolumn{2}{c}{1.0}                                   & phase\\
\hline
\multicolumn{4}{c}{\textit{Free, fitted parameter}} \\[0.1cm]
$R_{\rm p}/R_*$ & $0.1657^{+0.0072}_{-0.0078}$   & $0.146^{+0.023}_{-0.026}$\\
\hline
\multicolumn{4}{c}{\textit{Derived parameter}} \\[0.1cm]
$R_{\rm p}$     & $1.059^{+0.046}_{-0.050}$\,R$_{\rm opt}$     & $0.94^{+0.15}_{-0.17}$    & $R_{\rm opt}$\\
\hline
\end{tabular}
\end{table}

\subsection{UVM2 observation}
\begin{figure}
\centering
 \includegraphics[width=\columnwidth]{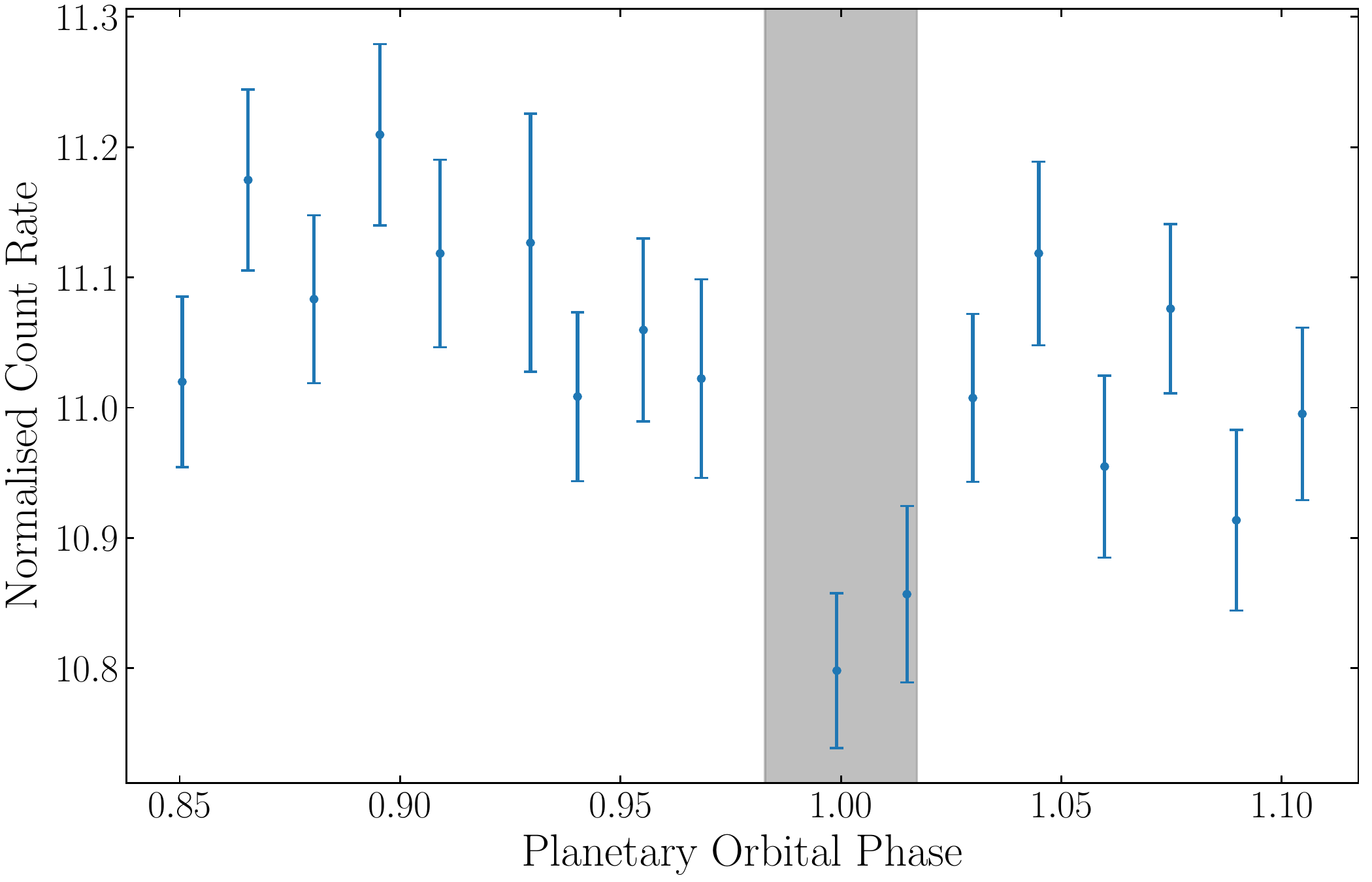}
 \caption{Binned OM UVM2 light curve for observation 1. This time series has not been corrected for out of transit trends.}
 \label{fig:UVM2raw}
\end{figure}

The observation on 17/18 April 2007 (observation 1 in Table~\ref{tab:Obs}) was taken using the UVM2 filter. The resulting light curve, displayed in Fig.~\ref{fig:UVM2raw}, showed evidence of a transit dip at the expected phase. We ran an MCMC fit to the time series data binned to 10\,s. We used a similar procedure to that outlined for the UVW2 data, with the same priors on $R_{\rm p, UVM2}/R_*$, $i$, $a/R_*$. The out of transit trend was again accounted for by multiplying the transit model by a quadratic in time. We derived limb darkening coefficients for the UVM2 filter using Limb Darkening Toolkit of u$_1$ = $0.0618\pm0.0250$ and u$_2$ = $0.0204\pm0.0250$, again using these values to place a Gaussian prior on the fit coefficients.

\begin{figure}
\centering
 \includegraphics[width=\columnwidth]{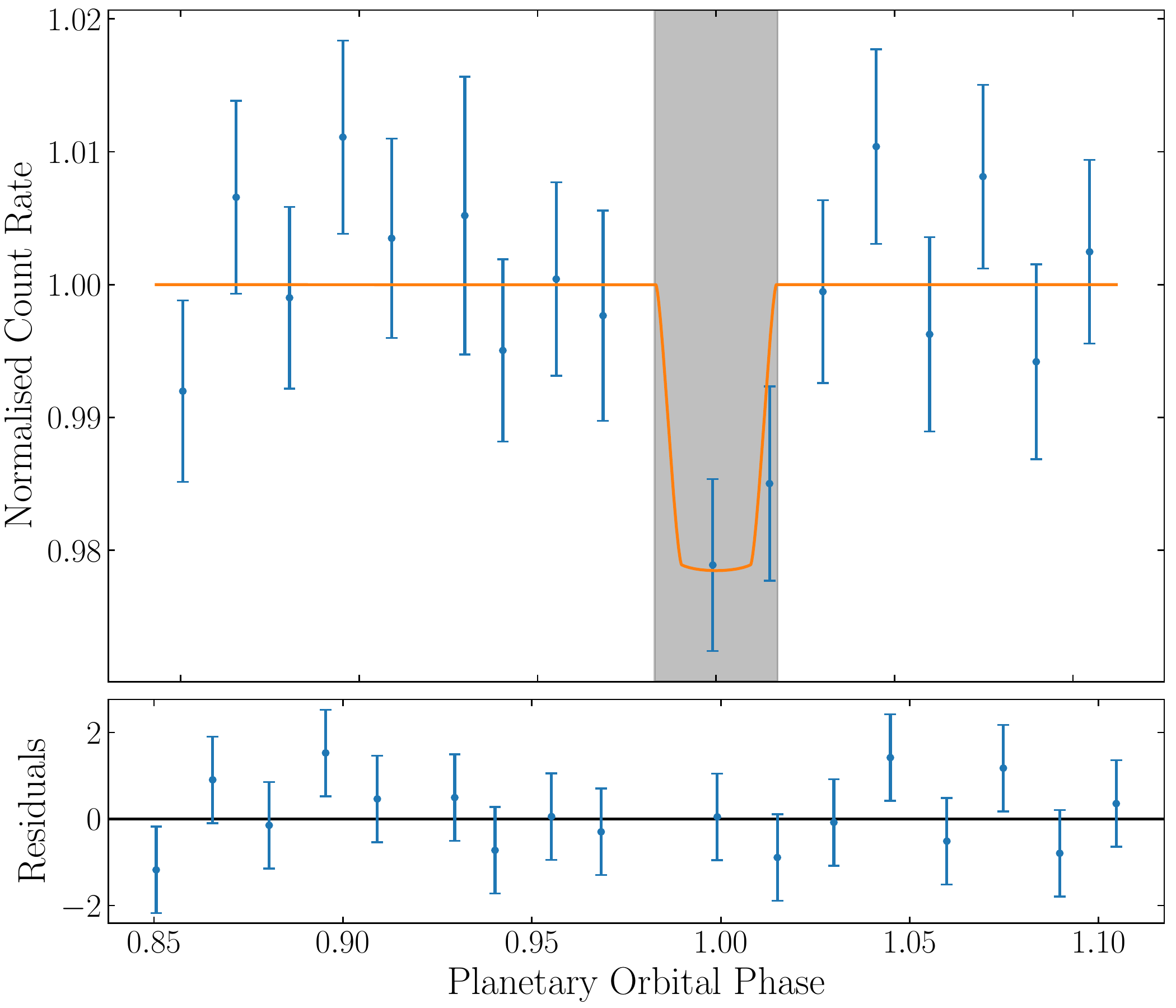}
 \caption{As Fig.~\ref{fig:om+model}, but for the single UVM2 light curve and associated fit.}
 \label{fig:UVM2fit}
\end{figure}

The light curve and the best fitting model from the MCMC are plotted in Fig.~\ref{fig:UVM2fit}, with the best fitting out of transit quadratic having been removed from the light curve. Parameter details for this MCMC run are again given in Table~\ref{tab:MCMC}. Our best fit value for the radius of the planet in this observation is $R_{\rm p, UVW2} = 0.94^{+0.15}_{-0.17}$\,R$_{\rm opt}$, and so consistent with the optically measured radius.

\section{Discussion}

\begin{figure}
\centering
 \includegraphics[width=\columnwidth]{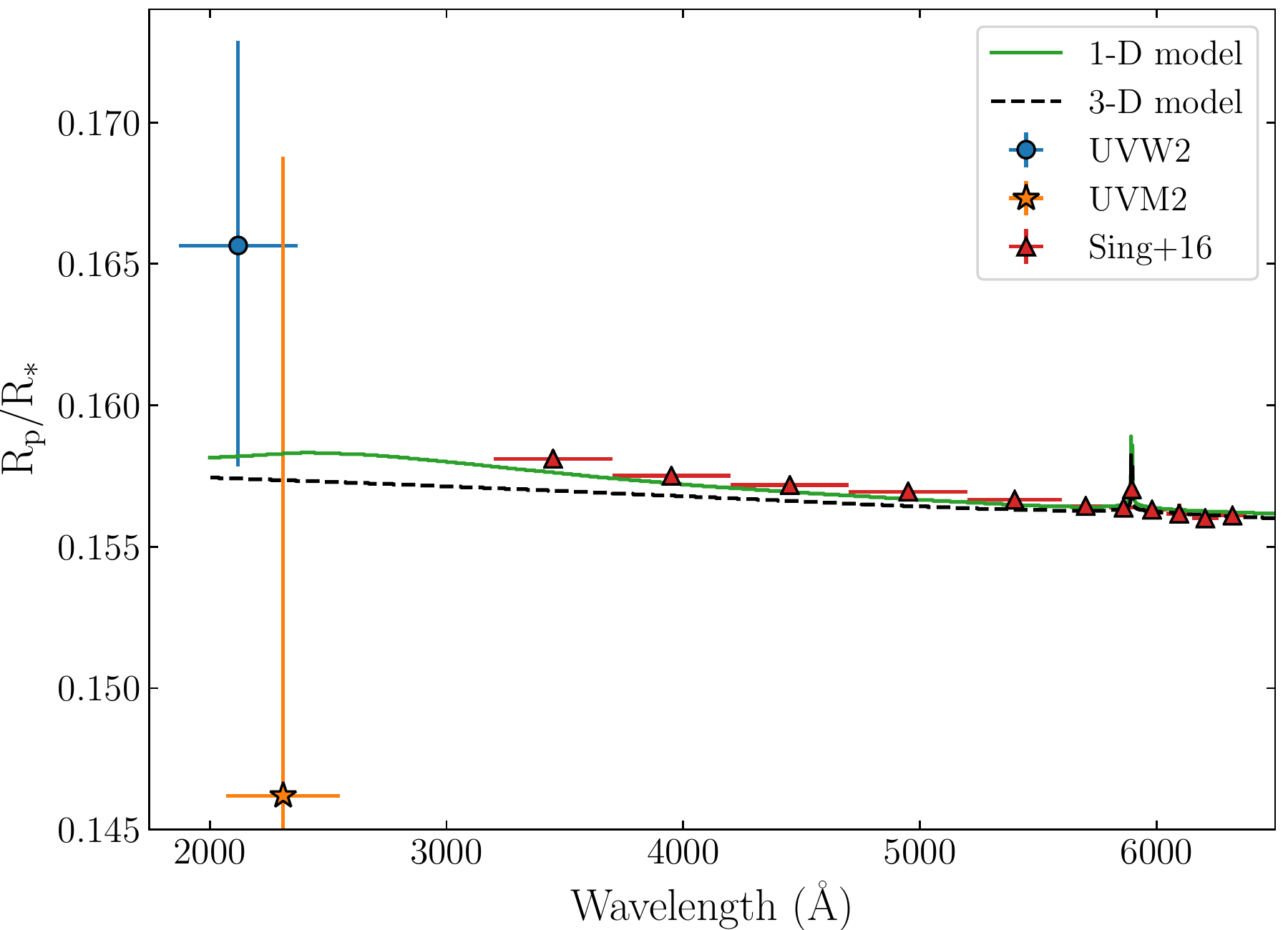}
 \caption{Transmission spectrum of HD\,189733b in the optical and NUV. The points associated with the transit measurements presented in this work are given with the blue circle (UVW2 filter), and orange star (UVM2). The red triangles are the data presented in \citet{Sing2016} (only that blueward of 6500\,\AA\ is displayed) - the errorbars on those points are too small to be seen on this scale. The green solid line is that expected by the 1-D model extended from the work of \citet{Lavvas2021}, and the black dashed line is that expected by the 3-D model extended from the work of \citet{Steinrueck2020}. The Roche lobe of the planet (3\,R$_{\rm opt}$, or $R_{\rm p}/R_* = 0.471$) is too far off the scale of the plot to be visible.}
 \label{fig:transSpec}
\end{figure}

The transmission spectrum of HD\,189733b exhibits a steep optical slope that may arise from either the presence of atmospheric aerosols or contamination from starspots \citep{Pont2008, Sing2011, Pont2013, McCullough2014}. In Fig.~\ref{fig:transSpec}, we plot a transmission spectrum of HD\,189733b. In addition to the two broadband filter measurements presented here, we include optical data bluewards of 6500\,\AA\,\citep{Pont2013,McCullough2014,Sing2016}. 
The single observation UVM2 point is not particularly informative as to whether the steep slope continues into the UV, given the large size of the error bar. 
One thing to note is that detailed simulations of aerosol formation in hot Jupiter atmospheres predict that transmission spectrum will eventually flatten at shorter wavelengths \citep{Powell2019}. Both the UVM2 and UVW2 data points are consistent with that scenario, although the errorbars are also too large to concretely prove or rule out enhanced absorption from other sources, such as FeII and MgII \citep{Turner2016,Salz2019,Lothringer2020}.

The measurements unquestionably rule out average opaque region sizes across these bandpasses similar to or exceeding the size of the Roche lobe (3\,R$_{\rm opt}$). Opaque region sizes measured at other wavelengths \citep[e.g. Ly\,$\alpha$:][]{LDE2012,Bourrier2020} have indicated there are detectable levels of escape of atmospheric material beyond this point. This and other such signals are typically observed in specific lines of interest and not in broadband measurements such as those presented here.


For comparison, in addition to the NUV and optical measurements, in Fig.~\ref{fig:transSpec} we also plot two recent transit spectra derived from simulations of the HD\,189733b atmosphere, extended to 2000~\AA. 
The 1-D model from \citet{Lavvas2021} includes disequilibrium chemistry and radiative feedback from photochemical hazes, which dominate the NUV opacity of the upper atmosphere, $0.1$~mbar and above. They found that hazes also work to heat the upper atmosphere above 1 mbar, which increases the effective scale height at short wavelengths. Utilising soot opacities, their model (green curve in Figure~\ref{fig:transSpec}) provides a good fit to the steep transmission curve for wavelengths less than 6000~\AA.  
The 3-D model from \citet{Steinrueck2020} features a GCM derived HD\,189733b transmission spectrum that includes the effects of a non-homogeneous distribution of photochemical hazes at the terminator. Assuming a uniform particle size of 3~nm and utilizing soot opacities, they find that enhancing the vertical mixing in the GCM (dashed black curve in Figure~\ref{fig:transSpec}) is required to approach the steep slope observed at short wavelengths, but their 3-D model fits the observed transmission spectrum at wavelengths longer than 6000~\AA\ better than the 1-D models.

The UVW2 measurement across 18 transits hints at a continuation of the steep blue slope into the NUV. However the uncertainties on our measurement mean that a flattening out of this feature, as seen in the 1-D model (see Fig.~\ref{fig:transSpec}) cannot be ruled out. We can however decisively rule out the average transmission region size across the UVW2 bandpass being similar in size to the Roche lobe, which is interesting in the context of previous studies of the neutral and singularly-ionised Fe and Mg lines a few hundred Angstroms either side of 2500\,\AA. Transits in these lines have previously been used to detect these species in the exospheres of exoplanets \citep{Fossati2010,Sing2019,Cubillos2020}. However, we cannot determine without higher resolution observations or further modelling whether there is no detectable exosphere Fe and Mg at these wavelengths, or if the broadness of the bandpasses used in this work have sufficiently washed out the deeper transits expected in those narrow lines.

\section{Conclusions}
We have observed the near-UV transit of prototypical hot Jupiter HD\,189733b in three broadband filters across twenty observations taken with the \textit{XMM-Newton} Optical Monitor. We successfully detected transits in two of these filters, UVW2 (18 observations) and UVM2 (1 observation), with the star proving too bright in the single UVW1 observation and saturating the camera. HD\,189733b is the third 
planet to have a near-UV transit detection by \textit{XMM-Newton} or \textit{Swift}.

With MCMC fits to the data using \textsc{batman} transit light curve models, we measured transit depths for UVW2 and UVM2 that are statistically consistent with the optically measured radius of the planet. We also show that there is no significant variation in $R_{\rm p,UVW2}/R_*$ across the 18 observations taken with UVW2 filter, within the measured uncertainties. The same conclusion was also reached when considering variation in $R_{\rm p,UVW2}/R_*$ between four defined epochs in Section~\ref{ssec:UVW2}.

Comparing with previous transmission spectra measurements taken with \textit{HST}, our measurement allows for either a continuation of the steep slope observed at optical wavelengths or a flattening off of this slope in the near-UV. Our results emphatically rule out the broadband transmission region size extending out close to or beyond the planet's Roche lobe, however such signals could still be possible in narrow wavelength regions around lines such of neutral and singly ionised Fe and Mg. Higher resolution observations are required to investigate the transit signal in those lines.

\section*{Acknowledgements}
PJW acknowledges support by STFC under consolidated grants  ST/P000495/1 and ST/T000406/1. DE has received funding from the European Research Council (ERC) under the European Union’s Horizon 2020 research and innovation programme (project {\sc Four Aces}; grant agreement No 724427). VB has received funding from the European Research Council (ERC) under the European Union's Horizon 2020 research and innovation programme (project {\sc Spice Dune}; grant agreement No 947634). This project has also been carried out in the frame of the National Centre for Competence in Research PlanetS supported by the Swiss National Science Foundation (SNSF).

\section*{Data Availability}
The \textit{XMM-Newton} data used in this work are publically available from the \textit{XMM-Newton} Science Archive: \url{http://nxsa.esac.esa.int/nxsa-web/#search}
 



\bibliographystyle{mnras}
\bibliography{ompaper} 






\bsp	
\label{lastpage}
\end{document}